# Implementing focal-plane phase masks optimized for real telescope apertures with SLM-based digital adaptive coronagraphy

**JONAS KÜHN,**[1,*] **POLYCHRONIS PATAPIS,**[1] **GARRETH RUANE,**[2] **AND XIN LU**[1]

[1]*Institute for Astronomy, ETH Zurich, Wolfgang-Pauli-Strasse 27, 8093 Zurich, Switzerland*
[2]*Department of Astronomy, California Institute of Technology, Pasadena, CA 91125, USA*
*\*jonas.kuehn@phys.ethz.ch*

**Abstract:** Direct imaging of exoplanets or circumstellar disk material requires extreme contrast at the $10^{-6}$ to $10^{-12}$ levels at < 100 mas angular separation from the star. Focal-plane mask (FPM) coronagraphic imaging has played a key role in this field, taking advantage of progress in Adaptive Optics on ground-based 8+m class telescopes. However, large telescope entrance pupils usually consist of complex, sometimes segmented, non-ideal apertures, which include a central obstruction for the secondary mirror and its support structure. In practice, this negatively impacts wavefront quality and coronagraphic performance, in terms of achievable contrast and inner working angle. Recent theoretical works on structured darkness have shown that solutions for FPM phase profiles, optimized for non-ideal apertures, can be numerically derived. Here we present and discuss a first experimental validation of this concept, using reflective liquid crystal spatial light modulators as adaptive FPM coronagraphs.

## 1. Introduction

Direct imaging is a promising exoplanet detection method, as it can potentially provide a contextual view of planetary-mass objects in their circumstellar environment, thus helping to improve our understanding about the underlying processes behind solar system and planet formation. Furthermore, the ability to separate the photons of a companion from its host star paves the way to high-resolution spectroscopy of atmospheres and time-resolved spectrophotometry. However, the level of contrast required to separate a planetary-mass companion from the glare of its host star varies between $10^{-6}$ to $10^{-12}$, depending on the planet mass, temperature, and host stellar type, and this at angular separations as small as a few diffraction beam widths (typically 10 to 100 mas for face-on Earth-like orbits at distances of 10 to 100 pc). This explains why only a dozen of directly imaged exoplanets have been found so far [1], essentially the thermal emission of young hot planetary bodies around very young (< 10 Myr) stars. In practice, reaching down to $10^{-7}$ contrast levels at small angular separation not requires exquisite wavefront control techniques to correct for the atmosphere and non-common path aberrations [2], but also the use of coronagraphic masks in an intermediate focal- [3, 4] or pupil-plane [5] to diffract the primary star point spread function (PSF) away from the scientific field-of-view of interest. Further, this has to be combined with advanced observing strategies like Angular Differential Imaging (ADI, see [6]) to enable PSF residuals subtraction, achieved through state-of-the art data reduction algorithm like LOCI [7] or PCA/KLIP [8]. High-contrast "planet imager" instruments meeting all these requirements have only been recently commissioned on 8+-m class telescopes [9-11], and are starting to deliver impressive first science [12, 13]. In the foreseeable future, better contrast will only be achievable from space observatories, e.g. like with WFIRST coronagraph [14], but ground-based observation will retain the angular resolution advantage, and even more so in the upcoming ELT-era. Extreme angular resolution at the diffraction limit of a 30- to 40-m telescope may for example open a discovery space around common M-stars, where the required contrast to image rocky planets are relaxed by two to three orders of magnitude, as compared to solar-type host stars [15]. However, this will require to be able to access angular scales between 1 and 2 $\lambda/D$, while current high-contrast instruments like SPHERE and GPI are limited to about 3 $\lambda/D$, and SCExAO to ~2 $\lambda/D$.

Although some coronagraphs are capable of reaching angular separations well within 1 $\lambda/D$ in presence of unaberrated wavefront [16, 17], all existing large telescope entrance pupils actually present some kind of obstruction and support structure to hold the secondary mirror (or even additional tertiary mirror, and sometimes other optics to reach Nasmyth platforms). Such more complex entrance pupil geometries detrimentally affect wavefront quality as compared to a flat-hat pupil profile, and this even for perfect correction provided by the AO system. In the case of high-contrast imaging, the drawbacks of having to deal with non-ideal obstructed pupils are numerous. First, achievable coronagraphic null depth is negatively affected, as all structures inside the entrance pupil will diffract light inside the high-contrast region [18]. Second, to deal with the first issue, a frequently-employed workaround solution is to oversize or reshape the pupil stop in the post-coronagraphic (Lyot) pupil plane, thus considerably affecting overall throughput, typically down to factor of 0.6 – 0.8. Third, the resulting reshaping of the PSF (e.g. "donut-shaped" PSF for a vortex coronagraph, see [18, 19]) distributes more unwanted starlight in the critical 1-2 $\lambda/D$ region, and it poses numerous hurdles for image registration processes during data reduction. Finally, the loss in low spatial frequencies caused by the central obscuration of the pupil, usually further pronounced by the oversizing of the Lyot pupil stop, results in a throughput hit at low angular separation, seriously affecting the coronagraphic inner working angle (IWA, the 50% throughput angular separation for a coronagraph): typically, a low-IWA coronagraph like a topographic charge-2 vector vortex, will see its IWA degrade from ~0.9 to ~2 $\lambda/D$ for realistic central obscuration to primary mirror ratios. Various strategies to deal with some of these key issues have been proposed and are at various degrees of readiness, among which are entrance pupil amplitude or phase apodization [20, 21], use of multiple coronagraphs in cascade [18, 19], or pupil-remapping strategies with dedicated optics [22]. Another recently suggested approach for FPM phase coronagraphy is to iteratively optimize the focal-plane 2-D coronagraphic phase profile in function of a given input pupil through numerical methods, a framework referred to as structured darkness [23, 24]. This can be done starting from a classical phase coronagraph profile, using Gerchberg-Saxton algorithms to iteratively minimize the starlight leakage in the high-contrast post-coronagraphic Lyot pupil plane, called the nodal area. Although the concept is highly attractive on paper, as it could in principle deal with any pupil shape, including support spiders but possibly also segmented pupils (which is directly relevant to any 10+-m class telescope), in practice the computed phase masks are next to impossible to manufacture using conventional approaches. Here we present a first experimental demonstration of this approach and a potential pathway for a real implementation at the telescope, using a new technological framework called Digital Adaptive Coronagraphy (DAC), based on the use of liquid crystal phase-only spatial light modulators (SLMs) as programmable pixelated focal-plane phase masks [25, 26].

## 2. Methods

### 2.1 Computing optimized 2-D coronagraphic phase masks with a GS algorithm

To derive the optimal focal-plane phase mask for a given entrance pupil, we used the same numerical framework as in [23], based on a Gerchberg-Saxton (GS) iterative loop. As illustrated in Fig. 1 for a charge-2 vortex coronagraphic phase profile [3] in the case of the VLT entrance pupil, which contains a secondary mirror central obscuration ($D_{sec}/D_{prim}$ ~ 0.15) and associated spider support structure, the main working principle consists of numerically imposing the suppression of the leakage in the nulled region of the post-coronagraphic Lyot pupil plane, the so-called nodal area. Then the field is Fourier-transformed back to the focal-plane, where corresponding phase correction is iteratively added to the original coronagraphic phase map.

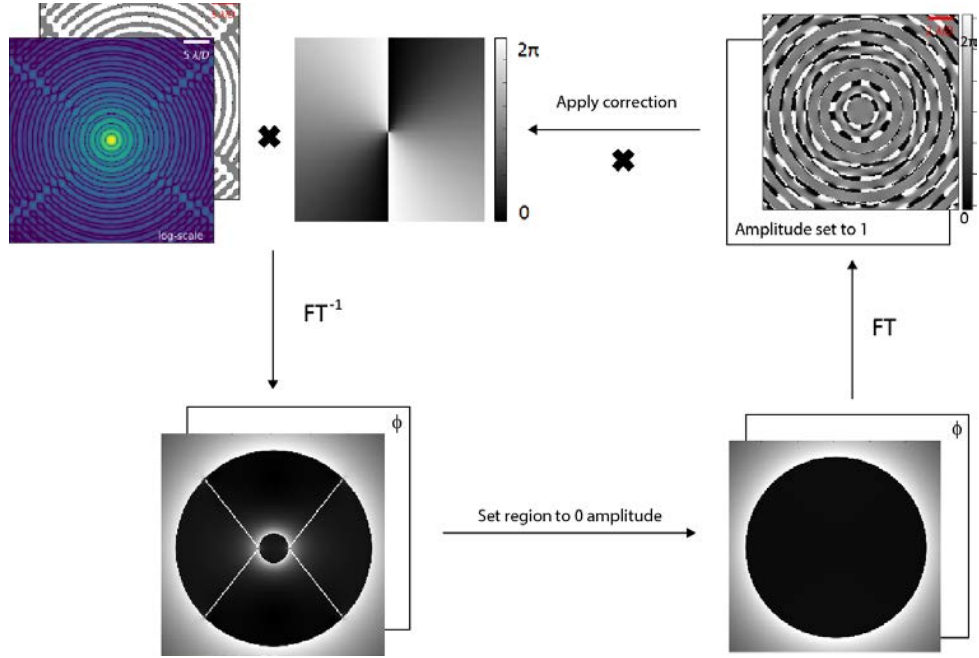

Fig. 1. Workflow chart of the modified GS algorithm used to the iteratively derive the focal plane coronagraphic phase masks dealing with non-ideal entrance pupil geometries.

In practice, the optimized phase masks are generated after about 400 iterations of the GS loop, with the simulation running on a 4096x4096 grid – about 10 times larger than the actual entrance pupil aperture - to ensure sufficient numerical accuracy when using Fast Fourier Transform (FFT) methods. This number of iterations has indeed been found to warrant a simulated coronagraphic null depth beyond $10^{-10}$. The corresponding computation time is then typically around 40 mn using Python on an Intel i7 processor (16GB of RAM).

## 2.2 Digital Adaptive Coronagraphy (DAC) high-contrast imaging optical testbed

In order to explore the feasibility and potential applicability of SLM-based Digital Adaptive Coronagraphy (DAC), a visible-range ($\lambda$ = 633 nm) coronagraphic testbed was built. The main particularity of using SLMs as adaptive focal-plane phase masks is that most competitive SLM displays (in terms of pixel size and fill factor, both key parameters for focal-plane usage) are based on the parallel-aligned nematic liquid crystal on silicon (PAN-LCOS) configuration, hence they should be used in an on-axis reflective configuration with linearly polarized impending light. This means that the coronagraphic focal-plane has to be implemented in a reflective arrangement, a configuration rarely seen on existing high-contrast instruments and testbeds. A scheme of the optical layout of this testbed is provided in Fig. 2, where the incident and departing beam on the PLUTO-series SLM (see §2.3 for details) are actually set at an angle of about +/- 3.5° versus the normal to the display surface.

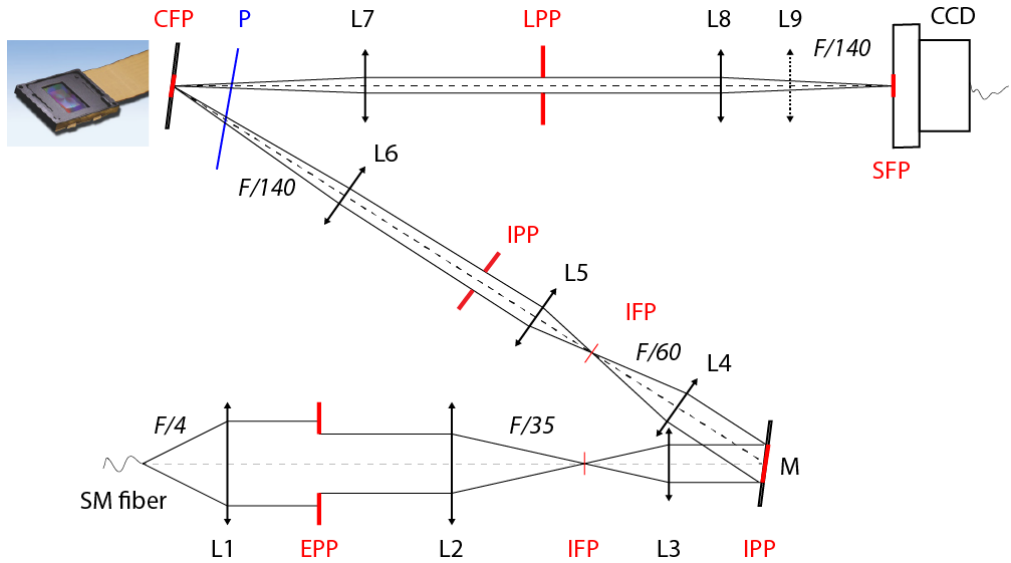

Fig. 2. Optical layout of the DAC high-contrast imaging testbed, not to scale and where the reflection angles are exaggerated to facilitate visualization (in practice: +/- 3.5° vs. the normal to the reflecting surface). L: aspheric lenses (100 to 250 mm focal lengths); M: fold mirror; P: linear polarizer; (E/I/L)PP: Entrance/Intermediate/Lyot pupil planes; (I/C/S)FP: Intermediate/Coronagraphic/Science focal planes. L8 and L9 are foldable lenses to enable focal-plane, respectively pupil-plane imaging.

Essentially, the DAC high-contrast imaging testbed of Fig. 2 consists of series of refractive plastic aspheric lenses (Edmunds Optics), placed in 4-f configurations to relay an unresolved point source to several intermediate pupil and focal-planes at various focal ratio. First, a 633-nm single-mode (SM) fiber-coupled laser diode source (Thorlabs) has its 5-μm output fiber tip acting as an unresolved point source (the "star"). Then the beam is collimated by a first lens (L1, f=150 mm) to a first arbitrary located entrance pupil plane (EPP), where a metallic mask can be installed as a mockup of a telescope aperture (typical diameter: 6-8 mm, about 4.5x smaller than the fiber tip diffracted beam waist, to ensure a flat-hat downstream beam profile). A pair of lenses (L2-3) then relay this pupil to a fold-mirror M (optionally a 12x12 Deformable Mirror from Boston Micromachines), which is then successively brought to focus twice, until the focal ratio enables at least 10 SLM pixels per $\lambda/D$ sampling in the reflective coronagraphic focal plane (CFP). After reflecting off the SLM adaptive phase mask at an incident angle of ~3.5°, the beam is then collimated by a lens (L7) to form a post-coronagraphic Lyot pupil plane (LPP), where the on-axis light diffracted by the phase mask can be blocked by a metallic mask, about 10% undersized as compared to the corresponding entrance pupil plane (EPP) aperture. Two foldable lenses (L8-9) finally perform either science focal-plane (SFP), or pupil-plane, imaging on the CCD camera (pco.pixelfly).

## 2.3 A PAN-LCOS Spatial Light Modulator as an Adaptive Focal-Plane Coronagraph

The functional details of liquid crystals (LC) SLM panels have been described in-depth elsewhere [27-30], and the study of their technical relevance and hardware limitations as adaptive coronagraphs for high-contrast imaging is currently still work-in-progress [26], thus not described in details herein. As a brief description, PAN-LCOS panels enable pixelated phase-only modulation of a linearly polarized incoming wavefront, by applying an electric field to each individual pixel cell to proportionally tilt the LC molecules along the z-axis, which will vary the amount of birefringence induced by the LC cell, and in turn modulate the phase for linearly polarized incoming light. When using SLM displays as programmable focal-plane coronagraphs, spatial sampling of the PSF, i.e. the number of addressable SLM pixels per λ/D units in the focal-plane, is obviously crucial for minimizing aliasing effects, and to be able to implement complex phase masks geometries. Combining this requirement with the need of high-fill factor properties, and phase-only response, led us to consider the PLUTO-series of reflective phase-only LCOS SLMs from Holoeye GmbH. Indeed, at first sight these panels meet several requirements deemed important for FPM coronagraphy, with pixel pitch below 10 μm, phase-only operation well beyond one wave retardance for linearly polarized incoming light, and fill factor in excess of 90%. The panel used in the present work is the PLUTO-VIS-014, whose main specifications at the λ=633 nm operating wavelength are listed in Table 1. Typically, at this wavelength, and with 8 μm-sized pixels, the PLUTO SLM device can achieve 10 pixels per λ/D units sampling in the focal-plane for a focal-ratio of about F/125, which is still relatively slow as compared to realistic instruments. However, given the PLUTO series has full compatibility with H-band operation (~ 1550 nm), the focal ratio requirement to achieve such a spatial sampling of the PSF at these near-infrared (NIR) wavelengths would decrease to a more acceptable ~F/60. Further, we note that the same company recently released a new "Gaea" 4K panel with 3.75 μm pixel pitch, which would further allow at least 10 pixels per λ/D PSF sampling in the NIR for focal-ratios as fast as F/25. Overall, this class of LCOS panels has potential for NIR phase coronagraphy applications at first look, although part of the technical readiness evaluation, including the present work, can be initially conducted in visible light.

In order to achieve accurate phase modulation in the focal-plane using this SLM panel, a phase response versus input 8-bits gray level calibration measurement was performed, using the same optical layout of Fig. 1. This was undertaken using an input pupil mask with three off-axis holes to illuminate different areas of the SLM display, and generate a "fringed PSF" on the imaging camera. Fourier-transform phase-domain analysis was then applied on each corresponding spatial frequency peak, while a step-by-step gray level modulation on only one illuminated region of the panel was performed. Differential phase measurement was finally computed on the three Fourier peaks, using the un-modulated pair of holes as reference signal for any setup instability. The final measured calibration curve for the effective phase modulation in function of input gray level video signal is plotted in Fig. 3.

Table 1. Holoeye PLUTO-VIS-014 main specifications for λ = 633 nm as used in this work

| Display panel type | Reflective PAN-LCOS |
|---|---|
| Pixel resolution | 1980 x 1080 |
| Pixel pitch | 8 μm |
| Fill factor | 93% |
| Addressing signal | 8 bits (256 grey levels) thru HDMI |
| Framerate | 60 Hz |
| Maximum retardance | 2.9 rad (using linearized response curve: 2.1 rad) |
| Phase stability | 0.07 rad rms |
| Reflectivity | > 65% |
| Incoming light SOP | Linearly polarized along display long axis |
| Acceptable off-axis incident angle | max. 5° vs. the normal to the panel surface |

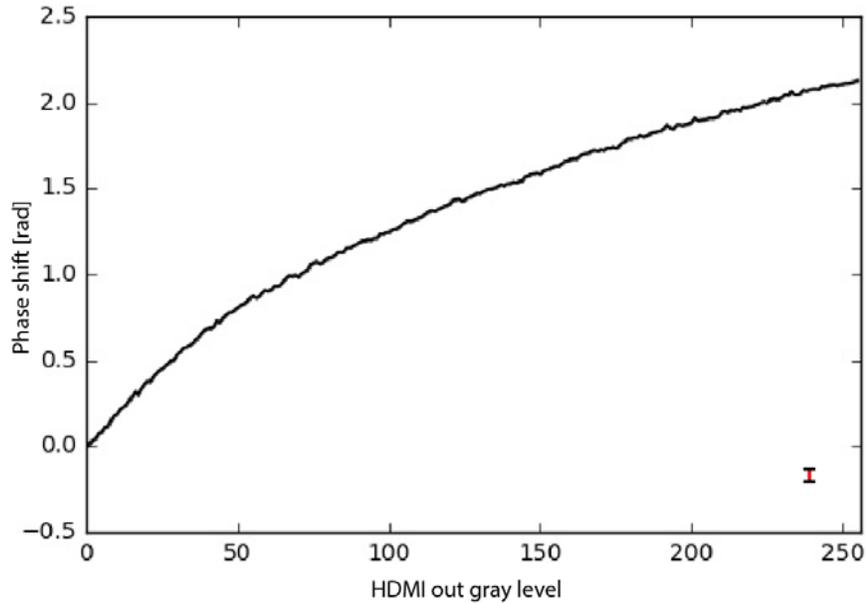

Fig. 3. Measured phase calibration curve for the PLUTO-VIS-014 SLM panel used in this work, in function of input gray level through the HDMI connection. The bottom right error bar indicates the typical panel phase rms, from the manufacturer specifications.

As indicated in Table 1, to achieve pure phase-only modulation of the incoming wavefront, only linearly polarized light should be illuminating the SLM panel. For a high-contrast application, on-axis leakage cannot be tolerated, which led us to retain an off-axis configuration with a slight ~3.5° angle versus the normal to the surface, theoretically within range of the manufacturer specification (Table 1). Indeed, the use of an on-axis scheme would have imposed to rely on some kind of polarized beamsplitter strategy to avoid mixing coronagraphically diffracted light with the incoming beam, where leakage due to A/R coating or polarization optics imperfections could ultimately compromise coronagraphic null depth. To ensure a linear state-of-polarization (SOP) with the best polarization contrast as possible (> 500:1 with our polarizer) on the display plane under this particular off-axis arrangement, we placed the polarizer (P) just in front of the panel, downstream of the last focusing lens (L6), as depicted in Fig. 1. This has been shown to improve performance as compared to placing the polarizer at the entrance pupil plane [EPP, Fig. 1], due to SOP degradation along the beam train, mostly caused by the 6 plastic aspheric lenses located upstream of the SLM panel. In addition, the polarizer also intercepts the SLM back-reflected beam, where it can perform as an analyzer to filter out SOP-altered components, originating from the off-axis geometry and, to a lesser extent, the non-collimated nature of the wavefront.

Coronagraphic performances on the bench using this SLM approach to program phase masks like vortices or n-quadrants are still modest in this preliminary experimental work, with typical monochromatic null depth in the order of $N \sim 2.5 \cdot 10^{-2}$ in absence of active correction (a 12x12 DM, installed at the fold mirror plane location on Fig. 2, has been shown to achieve $N \sim 1.5 \cdot 10^{-2}$ with phase-only correction). This "instrumental null depth" is thought to be dominated by residual amplitude wavefront errors, as well as zero-order leakage (unmodulated directly reflected light) from the SLM.

## 3. Results

*3.1 GS-retrieved optimized phase masks for VLT- and Palomar-type apertures*

We ran the GS-loop as described in §2.1 to iteratively retrieve focal-plane phase mask (FPM) solutions for the VLT and Palomar 200-inch telescope apertures, which span the existing range of secondary to primary mirror diameter ratio, as well as various geometrical orientations of the spider support structure. Numerical results obtained after 400 GS iterations are presented in Fig. 4 for the vortex coronagraph with topographic charge *n=2* [3], with the solutions dealing with all pupil features (central obscuration and spiders) and the ones optimized for the central obstruction only. Simulations dealing with all Palomar pupil features for topographic charge *n=4* vortex and four-quadrant phase mask (FQPM, see [4]) are also exposed in Fig. 5. Theoretical null depth *N* for each simulated configuration - i.e. the inverse of the total attenuation $A = N^{-1}$ obtained by computing the aperture photometric ratio of the non-coronagraphic PSF to the coronagraphic residuals - is indicated in each figure panel, and always exceeds $10^{-6}$ for the optimize-all solutions. The phase masks not dealing with the spiders logically exhibit worse nulls, but these could easily be improved to around the same order of magnitude using a dedicated spider cache in the Lyot stop, with limited impact on throughput. The resulting two-dimensional optimized phase profiles exhibit high spatial frequencies with concentric rings at λ/D scales, experimenting sharp π phase transitions. This is of course a concern for off-axis throughput (see §3.3), but also to first order make the manufacturing of such phase mask a daunting, if not impossible, task.

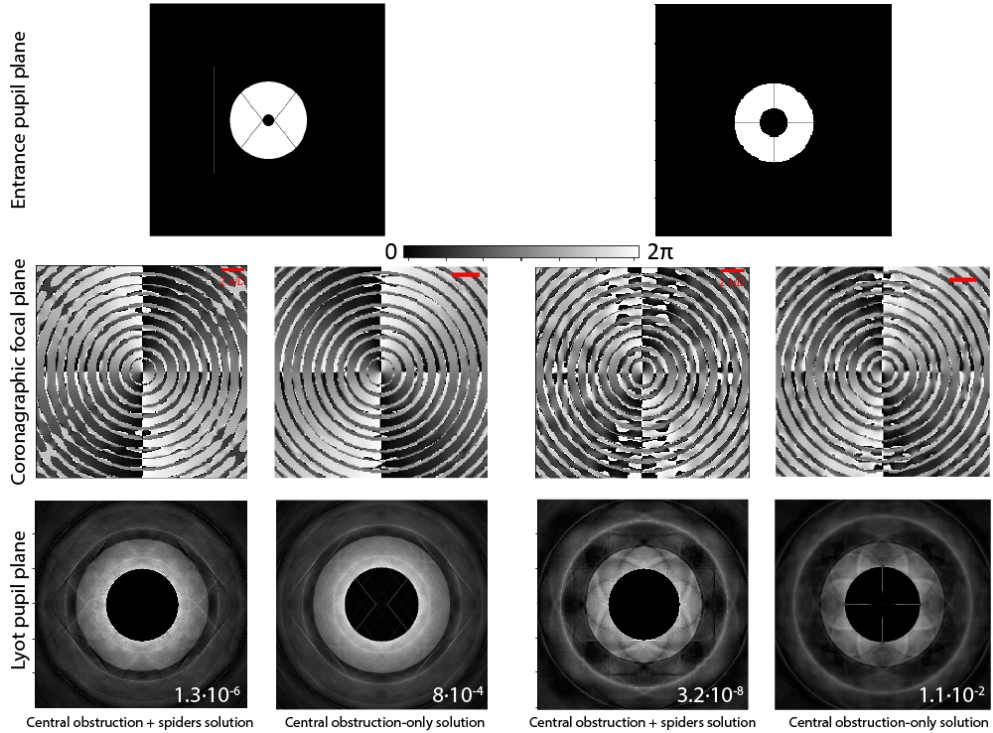

Fig. 4. Simulation of optimized masks derived from the vortex charge-2 initial conditions, for the VLT (left two columns) and Palomar (right two columns) entrance pupils. The optimized phase masks reject all the light inside the nodal area in the Lyot plane (exact null depth *N* numbers are overlaid in the bottom row Lyot pupil plane images). The red horizontal scale bars in the phase images corresponds to 2 λ/D in the focal plane.

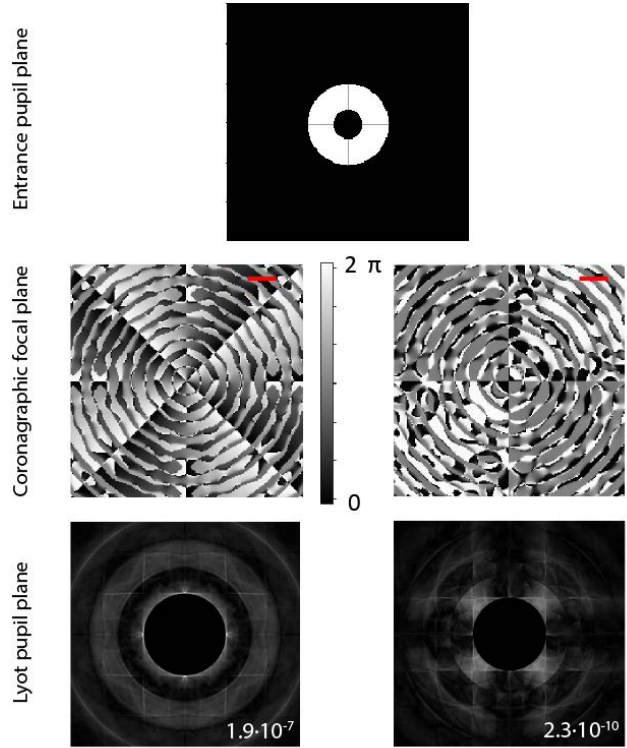

Fig. 5. Simulation and optimized masks for the Palomar entrance pupil, derived from the vortex charge-4 (left column) and FQPM (right column) initial conditions. Here the GS algorithm was set up so that the optimized phase masks deal with all the pupil features (secondary + spiders). The optimized phase masks reject all the light inside the nodal area in the Lyot plane (exact null depth $N$ numbers are overlaid in the bottom row Lyot pupil plane images). The red horizontal scale bars in the phase images corresponds to 2 $\lambda/D$ in the focal plane.

*3.2 Experimental implementation and initial results using the PLUTO SLM*

For pupil masks manufacturing reasons (laser cut Molybdenum 0.8 mm thick metallic masks fabricated in-house), we did limit ourselves to Palomar-shaped input pupils and Lyot stops here, as the spiders orientation and aspect ratio of the VLT pupil were out-of-reach of our in-house shop. Similarly, we had to oversize the Palomar 200-inch spiders as compared to real thickness of 0.25% to about 2% of the outer diameter (primary mirror), as the employed laser cut technology could not guarantee that spiders thinner than ~5% of their length would not break owing to their own weight. In all cases, we used 8-mm diameter entrance pupil masks to achieve F/140 focal ratio at the SLM panel location [Fig. 2], yielding ~ 11 pixels per $\lambda/D$ units spatial PSF sampling in the focal plane. The Lyot stop outer diameter is undersized by 10% relatively to the entrance pupil. An overview of the measured Lyot pupil plane, and focal plane light distributions in logarithmic scale (same scale for all PSF images), obtained by programming the SLM with the vortex charge-2 Palomar-derived solutions [Fig. 4, right] are depicted in Fig. 6, side by side with the theoretical simulation outputs of Fig. 4. Other SLM experimental results for the charge-4 vortex and the FQPM coronagraph are provided in Fig. 7. Table 2 lists the measured total null depth $N = 1/A$ measured with the camera imaging the pupil-plane, which in the specific cases of non-optimized FPMs can appear worse by a constant factor of a few as compared with the peak-to-peak attenuation $A_{ptp}$, owing to the PSF reshaping in the case of non-ideal pupils. However, at the telescope $A_{ptp}$ is usually easier to measure with high precision, and directly relates to the gain in integration time before starting to saturate the stellar residuals. Finally, a screen capture movie illustrating the SLM operations, iterating from a vortex charge-2 to its GS-optimized solutions, and ending up with the vortex charge-4 series of FPMs, is presented in the supplementary material (see Visualization 1).

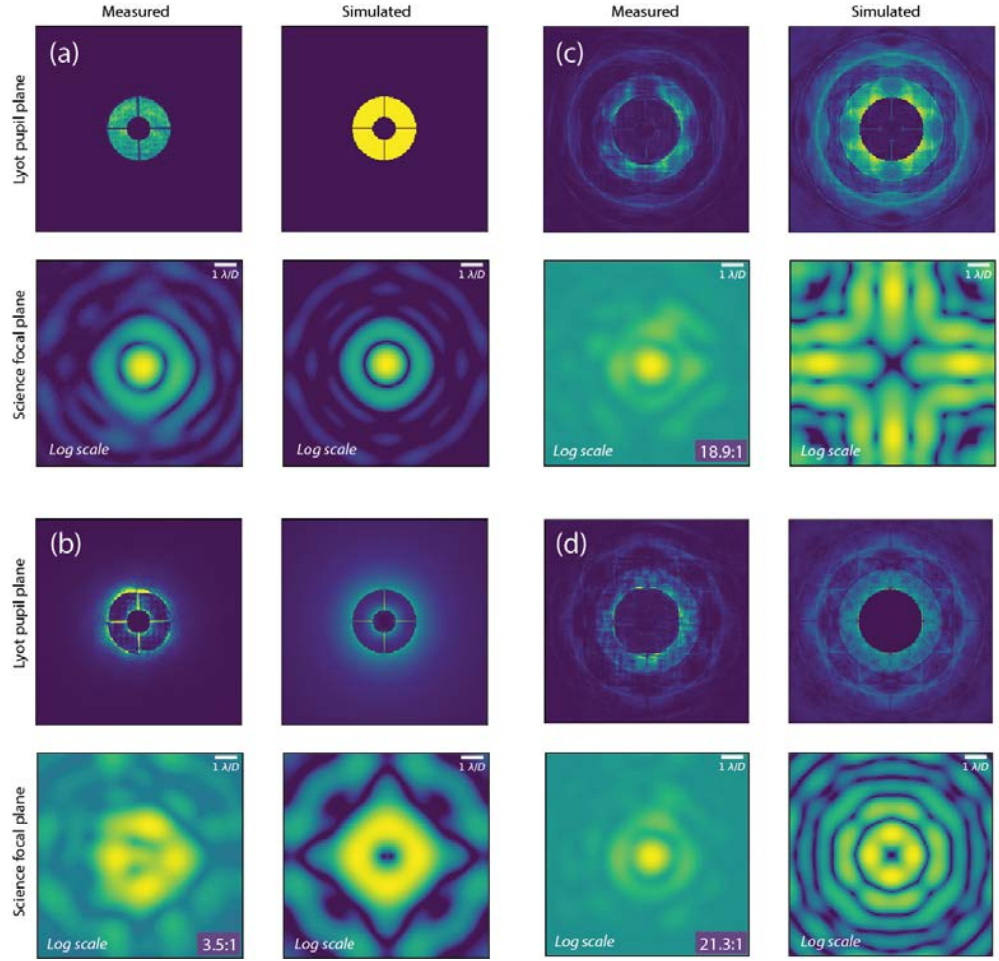

Fig. 6. Experimental results using SLM-based Digital Adaptive Coronagraphy (DAC) to implement GS-optimized phase masks derived from a vortex charge-2 initial focal-plane mask (FPM), to deal with the Palomar 200-inch pupil, and comparison with simulated brightness distributions. All Lyot pupil plane images are shown without the circular Lyot stop in place used for science focal plane recordings, where the measured total attenuation $A = 1/N$ is indicated. (a) Non-coronagraphic case, with a flatmap provided to the SLM; (b) Regular vortex charge-2 programmed on the SLM; (c) Same as (b) but for the GS-optimized FPM dealing with only the central obstruction; (d) Same as (b,c) but for the GS-optimized FPM dealing with all the Palomar pupil features (central obstruction and spiders).

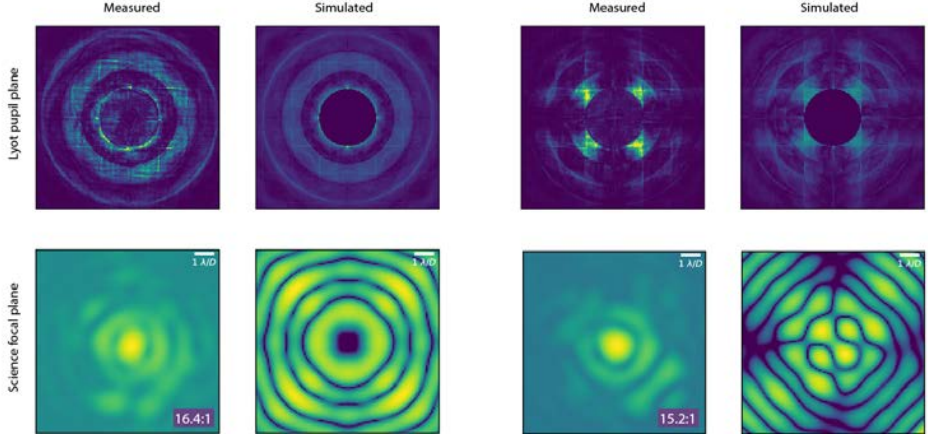

Fig. 7. Experimental results using SLM-based Digital Adaptive Coronagraphy (DAC) to implement GS-optimized phase masks of Fig. 5 derived from a vortex charge-4 (left) and a four-quadrant phase mask (FQPM, right) initial condition, to deal with the Palomar 200-inch pupil, and comparison with simulated brightness distributions as calculated in Fig. 5. All Lyot pupil plane images are shown without the circular Lyot stop in place used for science focal plane recordings, where the measured peak-to-peak attenuation $A_{ptp}$ is indicated.

Table 2. Experimentally measured total null depth $N = 1/A$ (measured in pupil-plane) for all FPM solutions dealing with the Palomar 200-inch telescope pupil.

| $\sigma_N = 5 \cdot 10^{-3}$ | Regular | Optimized for secondary only | Optimized for all pupil features |
|---|---|---|---|
| **Vortex $n=2$** | $28.4 \cdot 10^{-2}$ | $5.3 \cdot 10^{-2}$ | $4.7 \cdot 10^{-2}$ |
| **Vortex $n=4$** | $25.4 \cdot 10^{-2}$ | N/A | $6.1 \cdot 10^{-2}$ |
| **FQPM** | $31.4 \cdot 10^{-2}$ | N/A | $6.6 \cdot 10^{-2}$ |

In presence of a Palomar-shaped entrance pupil, the complex FPM experimental null depth numbers of Table 2 are in the order of $5 \cdot 10^{-2}$. This is about a factor two worse than instrumental nulls obtained with an unobstructed entrance pupil, when programing classical phase masks onto the SLM (see §2.3). Such a discrepancy could be explained by SLM platescale calibration errors (number of pixels per $\lambda/D$), potentially critical given the high spatial frequency content of some of the GS-retrieved FPM solutions, as well as low-order wavefront errors partially nullifying the gain of using the complex GS-solution FPMs. Devising precise platescale calibration procedures, and introducing some low-order wavefront errors within the GS computation loop, are therefore interesting avenues for investigation in the near-future.

*3.3 Raw contrast and throughput*

To quantify the gain in photometric dynamic range, it is common in high-contrast coronagraphic imaging to plot the azimuthally averaged numbers of counts in function of angular separation, normalized to the peak counts of the non-coronagraphic PSF. Such a so-called raw contrast plot is shown in Fig. 8 for the case of the charge-2 vortex coronagraph, and its two GS pupil-optimized derived solutions.

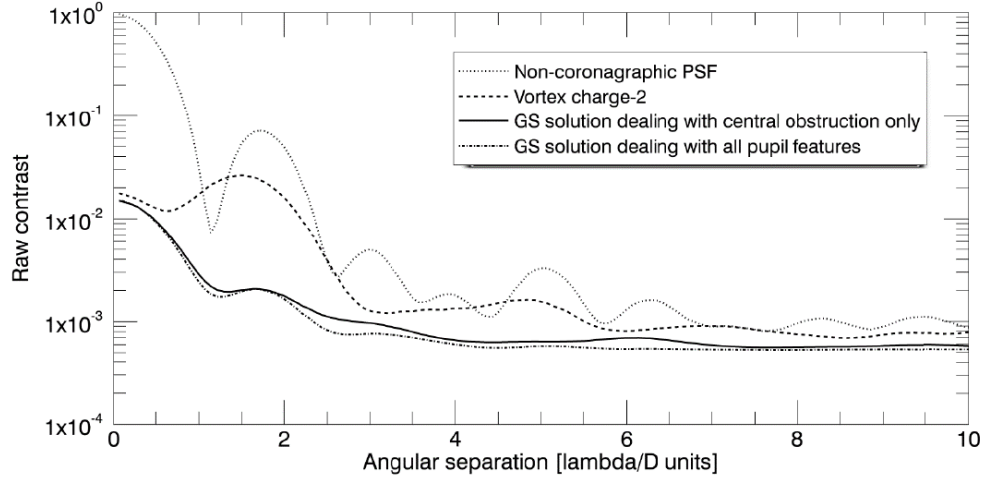

Fig. 8. Azimuthally-averaged raw contrast for the n=2 vortex coronagraph and the derived GS-optimized solutions, in presence of a Palomar 200-inch entrance pupil.

To correctly evaluate the real SNR gain provided by the optimized masks, however, off-axis throughput is a key parameter that has to be considered. Indeed, the GS-retrieved FPM solutions presented in section §3.1 clearly exhibit abrupt phase transition regions at each "ring location", that can be approximated as $\pi$ phase jumps. These can strongly affect the throughput of an off-axis point source, as compared to e.g. a smoother vortex phase helix. This is illustrated in Fig. 9, exposing the theoretical throughput response of the *n=2* vortex coronagraph FPM GS-solution dealing only with the 200-inch pupil central obstruction, which typically represents a "best case scenario" in terms of spatial phase derivative (least aggressive GS solution). To this end, Fig. 9 presents the two-dimensional throughput map for a diffraction-limited off-axis point source on a 40 x 40 $\lambda/D$ grid, the azimuthally averaged 1-D throughput plot in function of separation (given the high degree of azimuthal symmetry of the GS-solutions), and some key throughput order of magnitudes for a range of angular separations. Overall, the concern is less on the average off-axis throughput of ~ 0.15 (several other techniques dealing with obstructed pupils have throughput << 0.5), but on the spatial location dependency showing peak-to-peak variations of up to 0.3. The latter can seriously complicate accurate photometry retrieval of a potential bona-fide companion, and even more so for spatially extended sources like disks. Finally, the SNR gain $\Delta_{SNR}$ provided by the Fig. 9 FPM solution dealing with the 200-inch pupil central obscuration, as compared to using a classical charge-2 vortex mask in the exact same optical configuration, is plotted in Fig. 10 and calculated following Eq. (1):

$$\Delta_{SNR} = \frac{SNR_{GS-solution}}{SNR_{initial-FPM}} \propto \sqrt{\frac{T_{GS}}{C_{GS}} \frac{C_{init}}{T_{init}}} \qquad (1)$$

With $T_{GS}$, $T_{init}$ being the off-axis transmission characteristics of the GS-retrieved solution, respectively the initial FPM, i.e. the starting $0^{th}$ GS iteration (vortex charge-2 for this example), as plotted in Fig. 9, and $C_{GS}$, $C_{init}$ being the raw contrast (on-axis PSF residuals) achieved by the respective FPM, as plotted in Fig. 8.

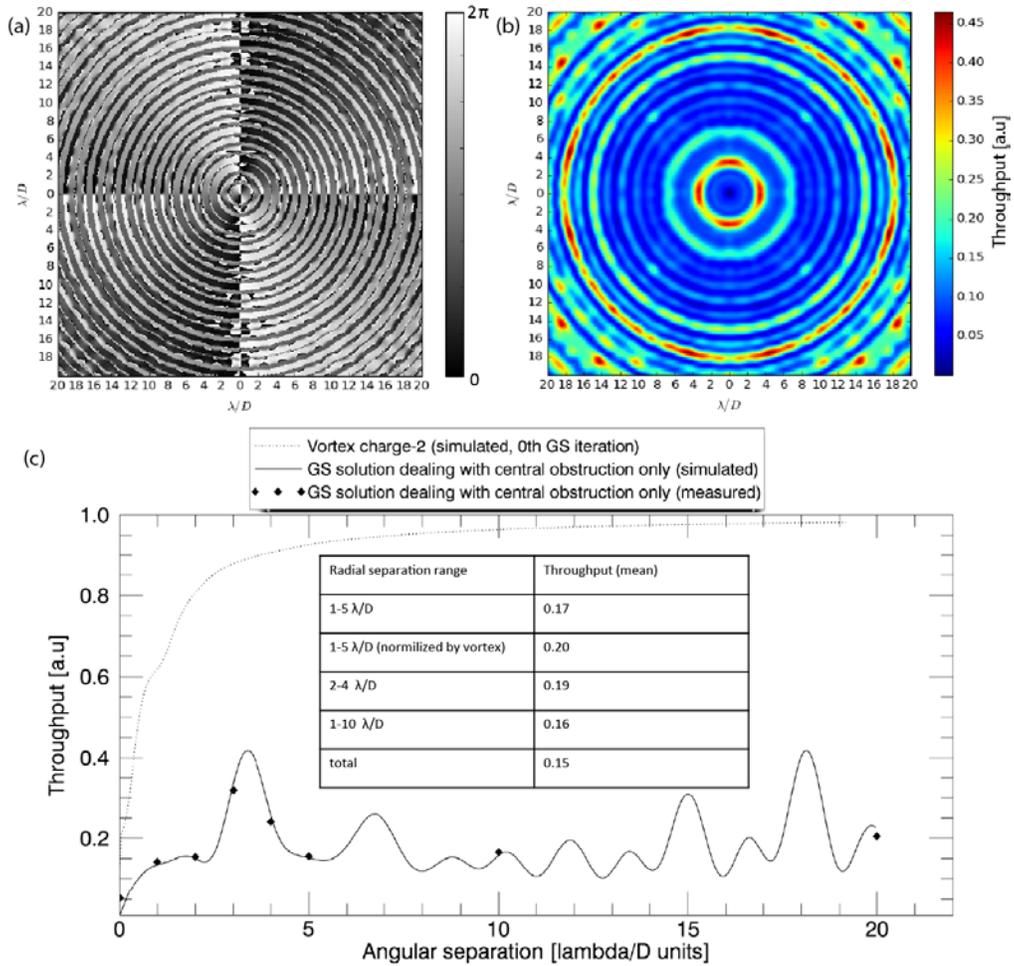

Fig. 9. Theoretical off-axis throughput characteristics of the GS-solution dealing only with the Palomar 200-inch central obstruction (not the spiders), with the vortex charge-2 as initial condition. (a) Numerical FPM solution after 400 GS iterations; (b) Two-dimensional throughput map of (a) for a diffraction-limited point source; (c) Azimuthally-averaged throughput of (b) in function of angular separation, compared with vortex charge-2.

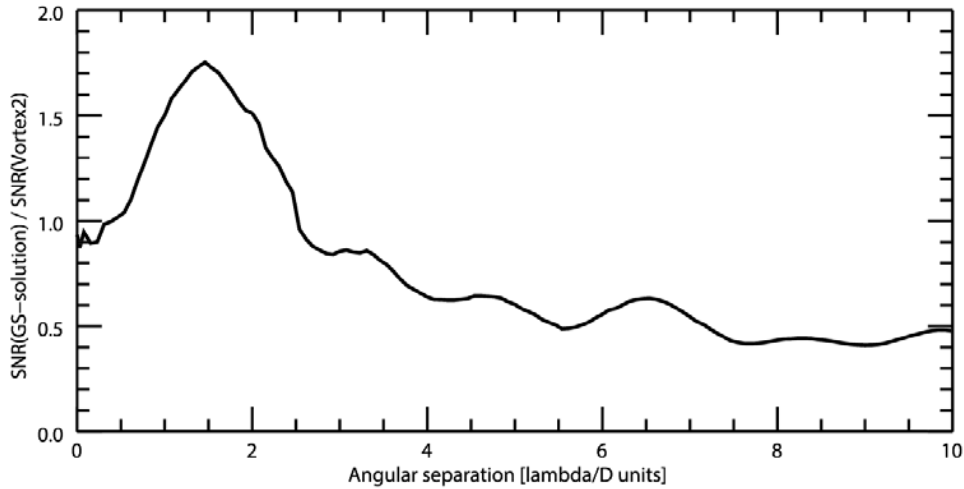

Fig. 10. SNR gain plot for a diffraction-limited point source in function angular separation, when the GS-solution of Fig. 9 is used as a coronagraphic FPM, as compared to a classical vortex charge-2 mask, located downstream of the Palomar 200-inch entrance pupil.

## 4. Discussions and perspectives

We were able to successfully implement complex numerical focal-plane phase masks solutions optimized for real telescope apertures, using a PAN-LCOS spatial light modulator (SLM) as an adaptive focal-plane phase mask coronagraph, working with monochromatic visible linearly polarized light. To first order, the post-coronagraphic Lyot pupil light distributions show high fidelity with the simulated results, enabling to create a homogeneously dark nodal area in the high-contrast region of the pupil, even in presence of a secondary mirror obstruction and its associated spiders support structure. The measured null depth quantities in the order of a few $10^{-2}$ are still modest, owing to residual low order aberrations on the bench (to be soon corrected with a deformable mirror) and potential systematic residual leakage from the SLM (still undergoing investigations), but they still mark a factor 4 to 5 improvement over classical phase masks. The off-axis throughput characteristics of the presented numerical phase mask solutions are far from ideal (~0.15 in average, but one should admittedly also factor in the overall SLM reflectivity of ~0.6 and the polarization loss of ~0.5 if one needs to estimate the total light efficiency), but the presented results have already brought a reduced parameter space between ~0.8 and 2.5 $\lambda/D$ into evidence. There, these masks should improve SNR as compared to their $0^{th}$-iteration counterparts, usually also used with oversized Lyot stop in the real world, and with typical light efficiencies < 0.75. In principle, any gain in experimental null depth should rapidly expand this "SNR gain region". We also note that it should be possible to add further constrains into the GS nodal area loop, for example to optimize throughput inside a specific area of the focal plane, or at a given angular separation. Overall, further works on deriving analytical solutions to these phase masks would be very valuable, and a preliminary analysis of the obtained FPM GS-solutions would tend to point towards a multiplication operation by a phase pattern matching the Fourier transform of an aperture-sized thin ring, pointing towards some kind of convolution operation in the pupil-plane responsible for ejecting the light outside the system aperture.

More generally, in light of the presented results, it seems clear that SLM-based Digital Adaptive Coronagraphy (DAC) is a promising approach to tackle several "niche" problematics in high-contrast imaging. With the inevitable complexification of telescope aperture shape in the wake of the ELTs, notably in terms of mirror segmentation, coronagraphic instruments will be operated even further away from the ideal wavefront conditions where they would shine their best. Part of the issues can be (and currently are) addressed through amplitude or phase apodization in the pupil plane, but any attempt to add complexity in the focal-plane phase function quickly present a challenge for most manufacturing techniques. Not only would a DAC framework contribute to deal with some of these aspects, but it could also do so in time-domain, as a pure "flip a switch" operation, hence potentially being able to re-configure the coronagraph to e.g. missing or defective mirror segments. The observational advantage could expand much further than exposed here, for example by enabling to change a vortex coronagraph topographic charge (i.e. its "aggressiveness") on the fly depending on weather conditions, trading inner-working angle for tip/tilt jitter insensitivity, or to be able to null several stars in the field during an ADI sequence, when observing compact binaries/triples [25]. Further work on near-infrared operation of SLM-based DAC is underway, and will also investigate bandwidth limitations, although the chromatic leakage term should, to first-order, be negligible down to a few $10^{-3}$ like for other phase coronagraphs. In this regard, we also note that SLM-based DAC retains a promising potential, by being able to implement pseudo-achromatic FPM solutions, like e.g. the dual-zone Roddier & Roddier family of FPMs [31]. Finally, in conjunction with a pupil-plane DM, a focal-plane SLM adaptive coronagraph could theoretically also act as a focal-plane active element of an "integral AO system", with the sole objective of optimizing contrast in real-time through a yet-to-be-developed close-loop system.

## Funding

JK and XL, as well as the entire DAC project, are funded through the Swiss National Science Foundation (SNSF) Ambizione grant #PZ00P2_154800.

## Acknowledgments

The authors would like to thank Prof. H. M. Schmid and Prof. M. R. Meyer for the research support since the DAC project started, as well as Dr. S. Daemgen, Dr. S. Quanz and Dr. A. Glauser for the motivating discussions and helpful contributions. We also thank Holoeye GmbH support staff for the timely and thorough answers to our questions on the SLM hardware.